\documentclass[a4paper,11pt]{article}
\usepackage{pos}

\title{Particle Identification at Future Colliders
}

\author*[a]{Roberto Preghenella}

\affiliation[a]{Istituto Nazionale di Fisica Nucleare, Sezione di Bologna (Italy)}

\emailAdd{roberto.preghenella@bo.infn.it}

\abstract{Particle identification (PID) remains a key ingredient of the physics programmes of future collider experiments. While traditional collider detectors rely on calorimetry, tracking and muon systems for particle classification, the identification of charged hadrons requires dedicated measurements of particle velocity through Cherenkov radiation, ionisation or time-of-flight techniques. Future facilities such as FCC-ee and the Electron-Ion Collider place demanding requirements on momentum coverage, detector integration and material budget, motivating the development of novel PID concepts. This contribution reviews several approaches currently under investigation, including compact Ring-Imaging Cherenkov detectors, cluster-counting drift chambers, timing-enhanced Cherenkov detectors and precision timing systems based on advanced silicon sensors. The role of emerging photodetector technologies and radiation-tolerance considerations for future collider environments is also discussed. This contribution summarises the main topics presented in the invited LHCP 2026 talk and focuses on selected representative examples rather than a comprehensive review of the field.}

\FullConference{
}

\begin{document}
\maketitle

\section{Introduction}

Particle identification (PID) is one of the fundamental capabilities of modern particle-physics experiments~\cite{Lippmann:2011bb}. Electrons, photons and muons can be efficiently identified through the characteristic signatures they leave in tracking detectors, calorimeters and muon systems. Charged hadrons, however, interact in a very similar manner inside a detector and cannot generally be distinguished solely on the basis of their energy deposition patterns.
The identification of charged pions, kaons and protons therefore requires an independent determination of the particle mass. Since the momentum is measured by the tracking system, the mass can be inferred through a measurement of the particle velocity. This principle forms the basis of most dedicated hadron-identification techniques employed in collider experiments, including Cherenkov imaging, time-of-flight measurements and ionisation-based methods~\cite{Lippmann:2011bb}.

Particle identification has played a central role in several major collider experiments. The flavour-physics programme of LHCb provides a particularly striking example of the impact of hadron identification on the reconstruction of complex final states and on the suppression of combinatorial backgrounds. Similar considerations apply to future collider facilities, where many key measurements rely on the capability to identify charged hadrons over broad momentum and angular ranges.
Future collider projects place new requirements on particle-identification systems. Future Higgs factories demand precise flavour tagging and exclusive reconstruction of hadronic final states while operating under stringent material-budget constraints~\cite{Wilkinson:2021ehf}. The Electron--Ion Collider requires identification of charged hadrons over a broad kinematic range in order to access the three-dimensional structure of nucleons and nuclei~\cite{AbdulKhalek:2021gbh}. At the same time, advances in photodetectors, precision timing and silicon technologies are opening new opportunities for detector concepts that were not feasible in previous generations of experiments.

This contribution provides an overview of the topics discussed in the invited talk ``Particle Identification at Future Colliders'' presented at LHCP 2026. Rather than attempting a comprehensive review of the field, it highlights a number of representative detector concepts and enabling technologies that illustrate current directions in the development of particle-identification systems for future collider experiments.

\section{Physics Drivers for Particle Identification}

\subsection{Future Higgs Factories}

\begin{figure}[t]
\centering
\includegraphics[width=\linewidth]{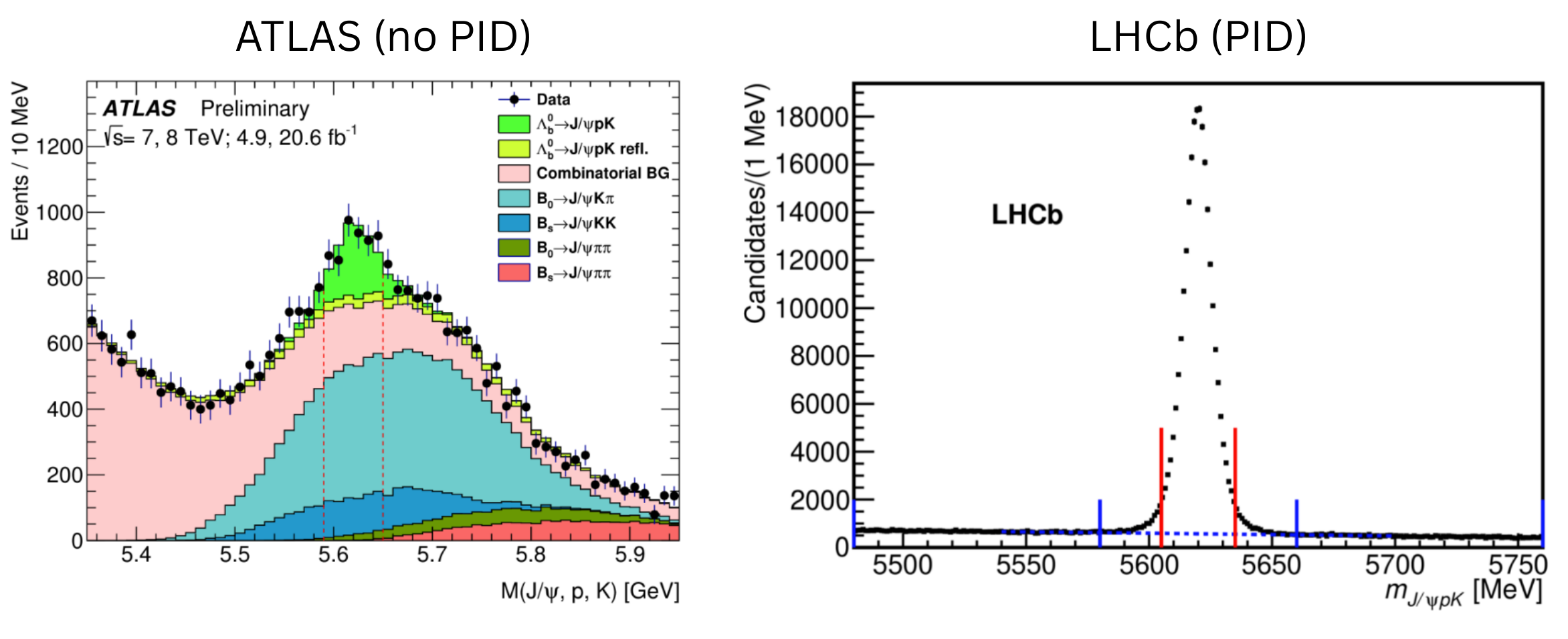}
\caption{
Invariant-mass spectra of reconstructed $\Lambda_b^0 \to J/\psi\,p\,K^-$ candidates used in pentaquark studies. The comparison between ATLAS, without dedicated hadron PID, and LHCb, with dedicated PID capabilities, illustrates the impact of particle identification on signal purity and background suppression. Adapted from Ref.~\cite{Wilkinson:2021ehf}.
}
\label{fig1}
\end{figure}

Future high-luminosity electron--positron colliders such as FCC-ee, CEPC, ILC and CLIC aim to perform precision studies of the Higgs boson, electroweak observables and flavour physics. While the clean environment of lepton colliders reduces many of the experimental challenges encountered at hadron colliders, efficient identification of charged hadrons remains crucial for a number of measurements.
Several flavour-physics observables rely on the reconstruction of exclusive hadronic final states and on the separation of pions, kaons and protons. Examples include measurements of CKM matrix elements through decays such as $B_s^0 \rightarrow D_s^\pm K^\mp$ , the study of heavy-flavour baryons through channels such as $\Lambda_b^0 \rightarrow J/\psi p K^-$ and searches for rare decays and exotic hadronic states~\cite{Wilkinson:2021ehf}. In these analyses, particle identification can dramatically improve the signal-to-background ratio and reduce combinatorial ambiguities.
As illustrated in Fig.~\ref{fig1}, dedicated hadron-identification capabilities can provide substantial improvements in reconstruction performance compared with detector systems relying solely on tracking and calorimetry. Such considerations motivate the inclusion of dedicated PID detectors in several future collider detector concepts.

\subsection{The Electron--Ion Collider}

\begin{figure}[t]
\centering
\includegraphics[width=0.7\linewidth]{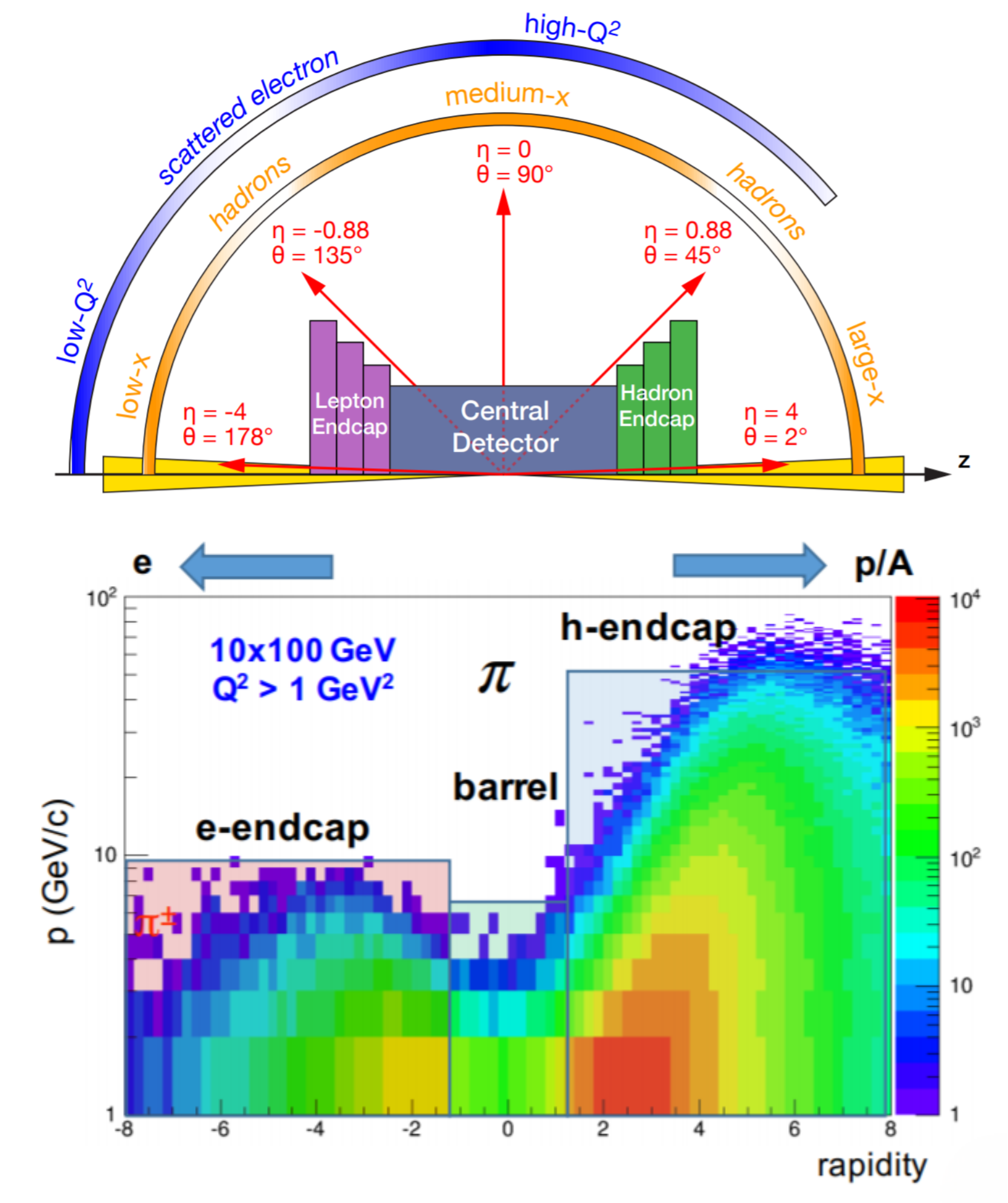}
\caption{
Particle-identification requirements at the Electron--Ion Collider. The broad momentum and rapidity coverage required for identified hadrons motivates the use of multiple complementary PID technologies. Adapted from Refs.~\cite{AbdulKhalek:2021gbh,Chatterjee:2024zrn}.
}
\label{fig2}
\end{figure}

The Electron--Ion Collider will investigate the structure of nucleons and nuclei through deep-inelastic scattering over a broad range of centre-of-mass energies and momentum transfers~\cite{AbdulKhalek:2021gbh}. Unlike Higgs factories, where particle identification is often used to improve the reconstruction of specific decay channels, identified hadrons constitute a primary observable of the EIC physics programme.
Semi-inclusive deep-inelastic scattering measurements require efficient separation of pions, kaons and protons in order to determine flavour-dependent parton-distribution functions, transverse-momentum-dependent distributions and fragmentation functions. Particle identification is also essential for studies of hadronisation in nuclear matter, strange-quark dynamics and exclusive processes.
Figure~\ref{fig2} illustrates the kinematic coverage relevant for identified hadrons in the EIC programme. No single PID technology can provide complete coverage of the required momentum and angular phase space, leading to detector concepts that combine multiple complementary techniques~\cite{Chatterjee:2024zrn}.

\section{Compact Cherenkov Detectors for Future Colliders}

Ring-Imaging Cherenkov (RICH) detectors remain one of the most powerful techniques for charged-hadron identification over broad momentum ranges. The measurement of the Cherenkov emission angle provides a direct determination of the particle velocity and, when combined with the momentum measured by the tracking system, allows the particle mass to be reconstructed.
Future collider experiments place increasingly stringent requirements on Cherenkov detectors. These include compact detector geometries, reduced material budgets, operation in magnetic fields, large active areas and compatibility with modern solid-state photodetectors. As a consequence, significant effort is being devoted to the development of new Cherenkov detector concepts that extend the capabilities of previous-generation systems.

\subsection{The ePIC dual-radiator RICH}

\begin{figure}[t]
\centering
\includegraphics[width=0.7\linewidth]{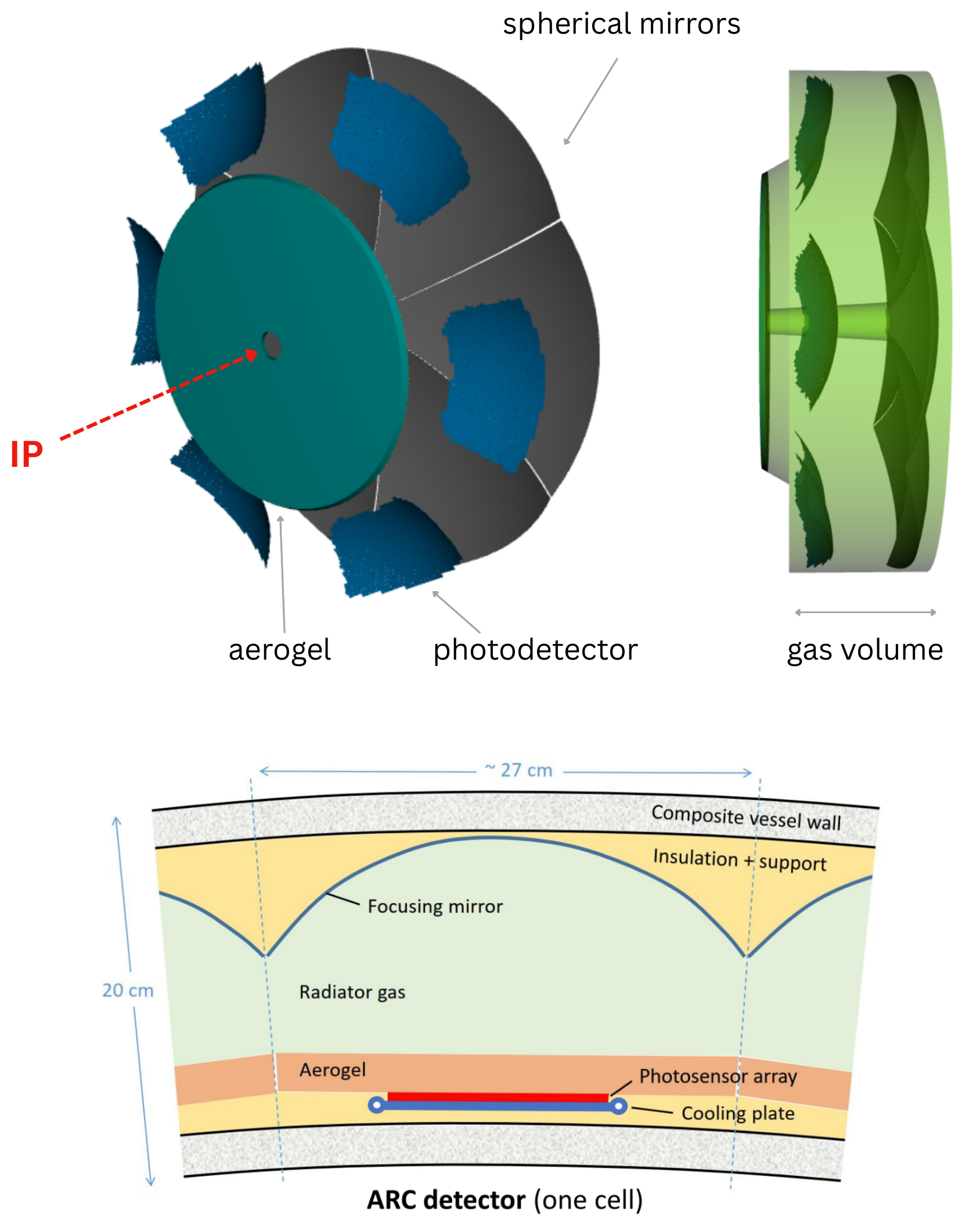}
\caption{
Examples of compact RICH concepts for future collider experiments. 
(top) The dual-radiator RICH (dRICH) detector developed for the ePIC experiment at the Electron--Ion Collider, combining aerogel and gas radiators for charged-hadron identification over a broad momentum range in the hadron-going direction. (bottom)
One cell of the ARC detector concept for future Higgs factories, where a compact dual-radiator optical system is integrated within a small radial envelope. Adapted from Refs.~\cite{Contalbrigo:2026bwu,Pezzulo:2026upv}.
}
\label{fig3}
\end{figure}

The hadron-going region of the ePIC detector at the Electron--Ion Collider requires charged-hadron identification over a momentum range extending from a few GeV/$c$ up to several tens of GeV/$c$. To satisfy this requirement, the ePIC detector employs a dual-radiator RICH (dRICH) detector combining aerogel and gas radiators within a compact optical system~\cite{Contalbrigo:2026bwu}.
The dRICH detector uses aerogel to provide particle identification at intermediate momenta, while a gaseous radiator extends the momentum reach towards higher values. Cherenkov photons are reflected by spherical mirrors and detected by large arrays of silicon photomultipliers (SiPMs), which offer excellent single-photon sensitivity, immunity to magnetic fields and high granularity~\cite{Rignanese:2024mco,Preghenella:2022rya}.
Figure~\ref{fig3} illustrates the detector concept. The use of two radiators allows continuous coverage across the momentum range required by the EIC physics programme while maintaining a compact detector geometry compatible with the overall ePIC layout.
The adoption of SiPM-based photon detection represents a significant evolution with respect to previous collider RICH detectors. The large active area required by the ePIC dRICH has motivated the development of dedicated cooling systems, low-noise front-end electronics and radiation-tolerance studies aimed at ensuring long-term detector operation~\cite{Preghenella:2023hgq}.

\subsection{The ARC detector concept}

A different set of constraints applies to future Higgs factories such as FCC-ee. In this case the experimental environment is comparatively clean, but detector concepts are often very compact and place stringent limits on radial space and material budget. Dedicated RICH detectors are therefore challenging to integrate, despite their excellent particle-identification performance.

The ARC concept, Array of RICH Cells, addresses this problem by replacing a single large RICH volume with an array of compact, optically independent cells~\cite{cardinale_2024_6entj-pmm10}. It has been proposed as a compact RICH detector concept for FCC-ee and future Higgs factories, in particular in the context of detector concepts based on all-silicon tracking. The target radial envelope is of the order of 20~cm, with a material budget below about $0.1\,X_0$.
Each ARC cell combines a dual-radiator configuration with compact focusing optics. A gaseous radiator, with C$_4$F$_{10}$ as baseline, provides high-momentum particle identification, while an aerogel radiator extends the separation power to lower momenta. The aerogel also provides thermal insulation for the photosensor plane. A spherical mirror focuses the Cherenkov light onto a highly granular photodetector array.
The ARC design exploits silicon photomultipliers as photon sensors. Their high photon-detection efficiency, fine granularity, magnetic-field compatibility and excellent timing capabilities make them attractive for compact collider RICH detectors. The relatively benign radiation environment of FCC-ee also reduces one of the main limitations associated with SiPM operation in Cherenkov applications.
Simulation studies indicate that the combination of aerogel and gas radiators can provide continuous $\pi/K$ separation at the level required for FCC-ee flavour physics, with a momentum reach extending up to about 40~GeV/$c$~\cite{Pezzulo:2026upv}. ARC therefore illustrates how classical RICH principles can be adapted to the geometrical constraints of future compact collider detectors.

\subsection{Timing-enhanced Cherenkov detectors}

\begin{figure}[t]
\centering
\includegraphics[width=0.7\linewidth]{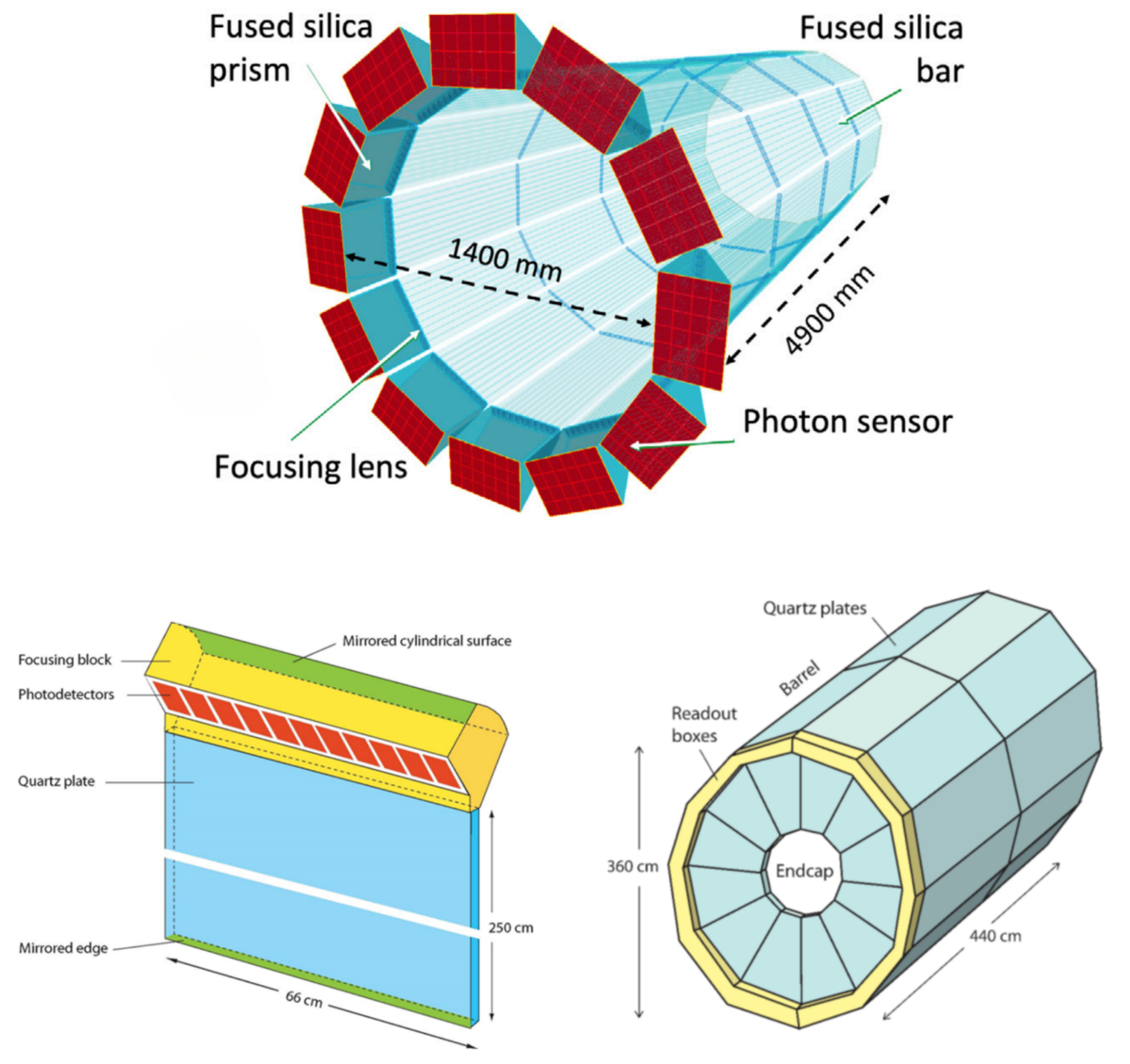}
\caption{
Examples of compact Cherenkov detectors for future collider experiments. (top) High-performance DIRC detector based on fused-silica radiators and precision photon imaging. (bottom) The TORCH concept, which combines Cherenkov imaging and precision timing to provide low-momentum hadron identification. Adapted from Refs.~\cite{Kalicy:2024jme,Charles:2010at}.
}
\label{fig4}
\end{figure}

An important trend in modern Cherenkov detector development is the increasing integration of precision timing information into the reconstruction process. Traditional DIRC (Detection of Internally Reflected Cherenkov light) detectors rely primarily on photon imaging. More recent concepts exploit both photon position and photon arrival time to improve reconstruction performance and reduce ambiguities.

The high-performance DIRC (hpDIRC) developed for ePIC represents an evolution of this approach. The detector employs fused-silica radiators, precision optics and highly segmented photon sensors to achieve excellent angular resolution while maintaining a compact geometry suitable for collider experiments~\cite{Kalicy:2024jme}.
A further development is represented by the TORCH concept, originally proposed for low-momentum hadron identification through precision time-of-flight measurements~\cite{Charles:2010at}. TORCH combines Cherenkov imaging and timing information by measuring the propagation time of photons inside thin quartz plates. The detector aims at timing resolutions of the order of a few tens of picoseconds per charged particle through the combination of many detected photons.
Figure~\ref{fig4} illustrates these two approaches. Together they exemplify the evolution of Cherenkov detectors from purely imaging devices towards systems in which timing information plays an increasingly important role in particle reconstruction.

\section{Cluster Counting and PID in Tracking Detectors}

\begin{figure}[t]
\centering
\includegraphics[width=0.7\linewidth]{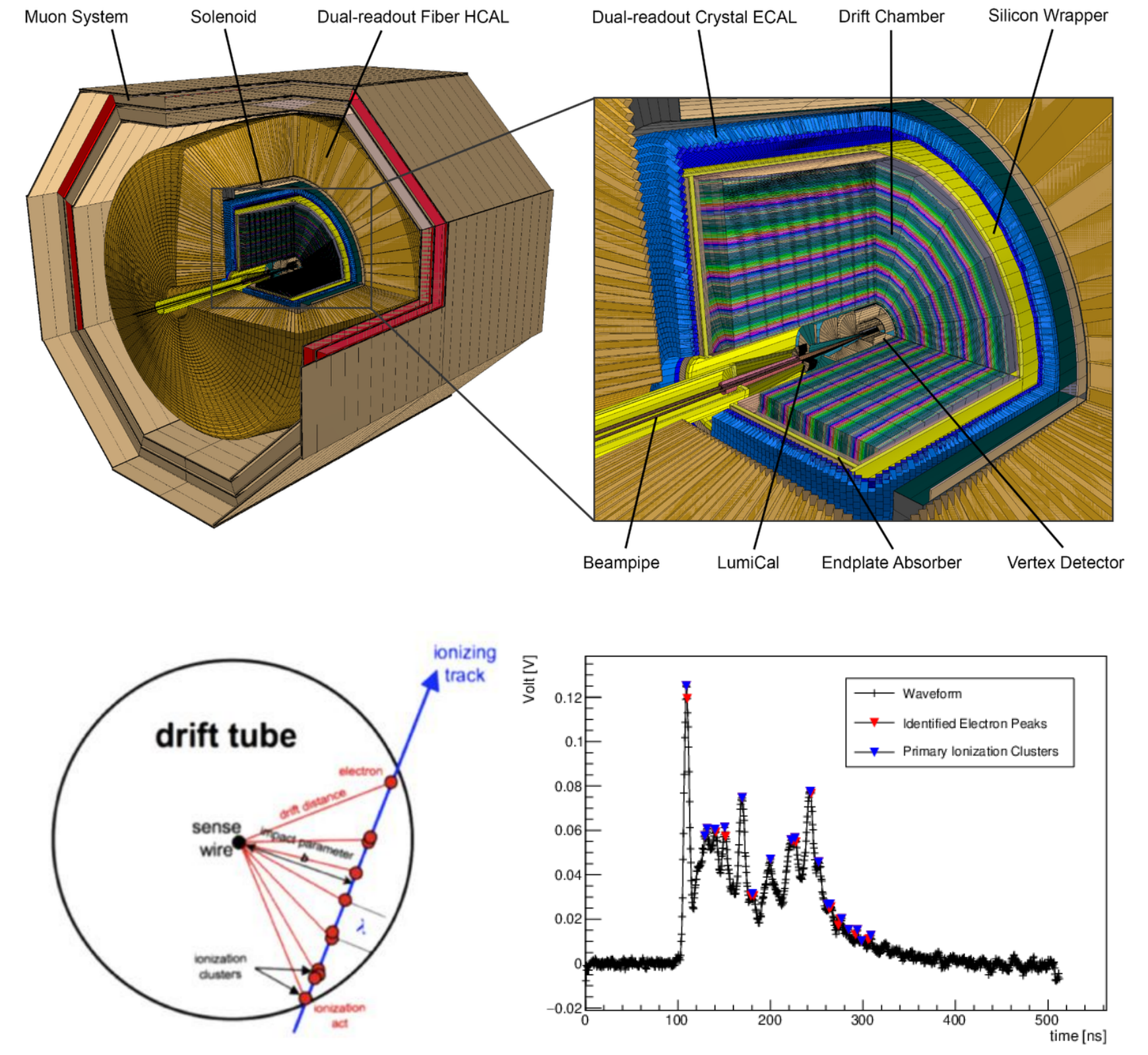}
\caption{
The IDEA detector concept and the cluster-counting technique. Particle identification is performed directly within the drift chamber through the measurement of the number of primary ionisation clusters rather than the total deposited energy. Adapted from Refs.~\cite{IDEAStudyGroup:2025gbt,Caputo:2022abo}.
}
\label{fig5}
\end{figure}

While Cherenkov detectors provide the highest performance over broad momentum ranges, alternative approaches seek to perform particle identification directly within the tracking detector. One of the most promising concepts in this direction is cluster counting.
Conventional gaseous tracking detectors exploit the average energy loss per unit length, $dE/dx$, for particle identification. The resolution of this technique is fundamentally limited by the large fluctuations associated with the ionisation process. Cluster counting instead measures the number of primary ionisation clusters produced along the particle trajectory, thereby reducing the impact of Landau fluctuations and improving separation power.

The IDEA detector concept~\cite{IDEAStudyGroup:2025gbt} for FCC-ee adopts this strategy through a large helium-based drift chamber capable of resolving individual ionisation clusters. Detailed studies have shown that cluster counting can significantly outperform conventional $dE/dx$ measurements, particularly in the momentum region relevant for flavour-physics applications.
Figure~\ref{fig5} illustrates both the IDEA detector concept and the principle of cluster counting. Analytical calculations predict excellent $\pi/K$ separation over a broad momentum range, with a narrow region around 1~GeV/$c$ where complementary timing information may be required.
More recently, detailed Garfield++ and Geant4 simulations have confirmed the expected performance up to momenta of approximately 20~GeV/$c$. In addition, dedicated beam tests have demonstrated the feasibility of efficient cluster reconstruction using fast waveform digitisation and advanced signal-processing algorithms. These results provide important experimental validation of the cluster-counting approach and support its consideration as a PID technique for future collider detectors~\cite{Caputo:2022abo}.

\section{Emerging Technologies and Future Challenges}

The requirements imposed by future collider experiments are driving rapid developments in detector technologies that extend well beyond traditional particle-identification systems. Advances in precision timing, silicon sensors and photodetectors are creating new opportunities for PID while simultaneously addressing challenges associated with detector integration, occupancy and radiation tolerance.

\subsection{Precision Timing}

Time-of-flight measurements have long been employed for particle identification at low and intermediate momenta. Historically, the achievable timing resolution limited their usefulness at high-energy collider experiments. Recent progress in detector technologies has dramatically changed this picture.
Over the last decade, silicon timing detectors based on the Low-Gain Avalanche Detector (LGAD) concept~\cite{Sadrozinski:2013nja} have demonstrated a steady improvement in performance. Early devices achieved timing resolutions approaching 30~ps for minimum-ionising particles~\cite{Lange:2017pxs}, while modern thin LGAD sensors routinely reach values close to 20~ps~\cite{Carnesecchi:2022tle}. Such performance opens the possibility of extending time-of-flight techniques into momentum regions that were previously inaccessible.

Precision timing is also becoming an increasingly important ingredient of Cherenkov detectors. Concepts such as TORCH and timing-enhanced DIRC systems combine spatial and temporal information to improve reconstruction performance and reduce ambiguities. In future collider experiments, timing information is expected to contribute not only to event reconstruction and pile-up mitigation but also directly to particle identification.

\subsection{Advanced Silicon Sensors}

The development of advanced silicon sensors is creating new opportunities for PID systems that combine precise timing with excellent spatial resolution~\cite{Preghenella:2020mxn}.
A particularly promising technology is represented by AC-coupled LGADs (AC-LGADs). Unlike conventional LGADs, which provide excellent timing performance but relatively modest spatial resolution, AC-LGADs distribute the induced signal over multiple readout electrodes. This approach allows the simultaneous reconstruction of the hit position and the signal arrival time with unprecedented precision.
Recent prototypes have demonstrated spatial resolutions at the level of a few micrometres while preserving timing resolutions comparable to those achieved by conventional LGAD devices. Such capabilities make AC-LGADs attractive candidates for future tracking and timing systems where precise measurements of both space and time are required~\cite{Apresyan:2020ipp}.

At the same time, developments in SPAD-based technologies and silicon photomultipliers continue to push the limits of single-photon detection. Originally developed for photon counting applications, SPAD arrays have also demonstrated excellent timing performance for charged-particle detection~\cite{Carnesecchi:2023dfq}, illustrating the increasing convergence between tracking, timing and photodetection technologies.

\subsection{Radiation Environment}

The operating environment plays a crucial role in determining the suitability of a given PID technology. Future collider projects span a remarkably broad range of radiation conditions.
Future Higgs factories such as FCC-ee, CEPC and the ILC are expected to operate in comparatively benign radiation environments. Detector optimisation can therefore focus primarily on performance, material budget and integration constraints.
The situation changes significantly for hadron colliders and muon colliders. FCC-hh will operate at centre-of-mass energies approaching 100~TeV and at unprecedented luminosities, leading to very large particle fluxes and radiation levels throughout the detector volume. Technologies developed for future PID systems must therefore combine excellent performance with long-term radiation tolerance.
Radiation effects are particularly relevant for silicon photomultipliers~~\cite{Garutti:2018hfu}. Increased dark-count rates, changes in gain and noise characteristics, and the need for low-temperature operation become important design considerations for large-area photon-detection systems. These issues have motivated extensive irradiation campaigns and dedicated studies aimed at understanding the long-term behaviour of modern photodetectors.

\subsection{Muon Collider Considerations}

Muon colliders present a unique detector environment among future collider proposals~\cite{Accettura:2023ked}. Unlike proton colliders, the primary collisions occur in a relatively clean lepton-collider environment. However, the finite lifetime of the circulating muon beams generates intense beam-induced backgrounds.
Electrons produced by muon decays interact with accelerator components and produce large fluxes of secondary particles that can enter the detector volume. In addition, electromagnetic interactions between the colliding beams generate large numbers of low-energy electron--positron pairs through two-photon processes. These backgrounds can contribute significantly to detector occupancy and radiation dose, particularly in the innermost detector regions~\cite{MAIA:2025hzm,Andreetto:2025mrd}.
As a consequence, particle-identification systems for muon colliders must be designed with particular attention to timing performance, occupancy rejection and radiation tolerance. The combination of precision timing and highly segmented detectors is expected to play a central role in mitigating beam-induced backgrounds while preserving PID performance.

\section{Summary and Perspectives}

Particle identification remains a key ingredient of the physics programmes of future collider experiments. Future Higgs factories and the Electron--Ion Collider place demanding and complementary requirements on detector systems, motivating the development of novel approaches to hadron identification.

Compact Cherenkov detectors such as the ePIC dRICH and hpDIRC demonstrate how modern photodetectors and optical designs can extend the capabilities of traditional RICH and DIRC technologies. At the same time, concepts such as TORCH illustrate the growing importance of precision timing in particle identification. An alternative direction is represented by cluster-counting drift chambers, which seek to perform PID directly within the tracking detector through measurements of primary ionisation clusters.
The rapid evolution of silicon technologies, including LGADs and AC-LGADs, is opening new opportunities for detector systems capable of combining precise timing and spatial measurements. In parallel, advances in silicon photomultipliers and SPAD-based sensors continue to expand the possibilities for photon detection in future experiments.
Finally, the increasingly diverse environments foreseen for future colliders, ranging from the relatively benign conditions of Higgs factories to the challenging backgrounds expected at FCC-hh and muon colliders, require careful consideration of radiation tolerance and detector robustness.

Future collider physics will continue to rely on advances in particle-identification systems and technologies. The developments reviewed in this article illustrate how innovation in detector concepts, photodetectors, timing systems and silicon sensors is shaping the next generation of PID capabilities.

\end{document}